\PassOptionsToPackage{sort&compress}{natbib}
\documentclass[aps,preprintnumbers,superscriptaddress,showpacs]{revtex4}
\usepackage{epsfig}
\usepackage{psfrag}
\usepackage{amsfonts}
\usepackage{graphicx}
\usepackage{dcolumn}
\usepackage{bm}
\usepackage{amsmath}
\usepackage{relsize}

\begin{document}

\title{Spin alignment of vector mesons in a plasma at finite density}

\author{Fei Wang}
\affiliation{School of Mathematics and Physics, China University
of Geosciences (Wuhan), Wuhan 430074, China}
\author{Zi-qiang Zhang}
\email{zhangzq@cug.edu.cn} \affiliation{School of Mathematics and
Physics, China University of Geosciences (Wuhan), Wuhan 430074,
China}

\begin{abstract}
Holographic predictions for vector meson spin alignment at finite
density have been limited by how the chemical potential is
implemented. Existing models use a baryon chemical potential and
report that the alignment is insensitive to it, while zero-density
benchmarks do not address the geometric deformation. We address
this limitation by using a charged AdS black-hole background with
a quark chemical potential and solving the coupled bulk vector
equations with infalling horizon conditions. The zero-density
limit reproduces previous results of Sheng et al. [Phys. Rev. D
110, 056047 (2024)]. At \(T=0.15~\mathrm{GeV}\), the \(J/\psi\)
spectral function has a resonance peak at \(\mu=0\) that lowers
and broadens with increasing \(\mu\), whereas the \(\phi\)
spectral function shows no pronounced peak. In the helicity frame,
\(\rho_{00}\) deviates positively from \(1/3\) for the \(J/\psi\)
and negatively for the \(\phi\). The density response is more
visible in our setup than the weak baryon chemical potential
sensitivity reported in a four-flavor soft-wall model. This
difference illustrates the model dependence of the density
response: it is controlled by the chemical potential
normalization. We identify the presence of the resonance peak and
the signs of the helicity deviations as robust qualitative
features, and treat the numerical shifts of \(\rho_{00}\) as
model-dependent estimates.
\end{abstract}


\maketitle

\section{Introduction}

The quark-gluon plasma (QGP) is a deconfined phase of quarks and
gluons that exists only under extreme temperatures and
densities~\cite{Bass_JPG1999,Scherer_PPNP1999,Shuryak_PPNP2009}.
Relativistic heavy-ion collisions provide the primary experimental
means to create this state in the laboratory. Yet the QGP is
exceedingly short-lived and cannot be observed directly. All
information about it must be extracted from the spectra and
correlations of the final-state particles. This inversion problem
makes the quantitative extraction of fundamental properties, such
as the equation of state, transport coefficients, and microscopic
structure, particularly challenging.

The sensitivity of $J/\psi$ to the surrounding medium has long
motivated its use as a QGP probe~\cite{Matsui_PLB1986}. Over the
past three decades, gauge/gravity
duality~\cite{Maldacena_ATMP1998,Gubser_PLB1998,Witten_ATMP1998}
has matured into a powerful nonperturbative tool for studying
heavy quarkonia in hot strongly coupled matter. Holographic
studies have examined meson spectral
functions~\cite{Zhao_EPJC2022,Zhao_EPJC2023,Braga_PLB2017,Braga_PLB2018,Mamani_PRD2022,Braga_PRD2023,Zhu_CPC2025,Cao_JHEP2021,Zhu_EPJC2024,Toniato_PRD2025},
thermal
widths~\cite{Zhao_EPJA2020,Feng_PRD2020,Braga_PRD2016,Finazzo_JHEP2013},
configuration
entropy~\cite{Zhao_PLB2023,Braga_PLB2023,Braga_PRD2022,Braga_PLB2021b,Braga_PLB2021a},
heavy-quark
potentials~\cite{Chen_JHEP2024,Zhou_PRD2020a,Zhou_PRD2020b,Zhang_NPB2023,Tahery_JHEP2023,Zhang_JHEP2011,Chu_JHEP2009,Hou_JHEP2008},
and other properties of hot and dense
media~\cite{Hou_PLB2021,Chen_PRD2018,Chen_EPJC2024,Zhu_EPJC2020,Chen_CPC2022,Wu_CPC2022,Li_CPC2021,Zhu_arXiv2023}.
This body of work has established holographic methods as a useful
framework for describing quarkonium in hot and dense
environments~\cite{CasalderreySolana_arXiv2011,DeWolfe_PPNP2014,Sadeghi_AHEP2013}.

Measurements by the STAR and ALICE Collaborations have reported
spin alignment of $\phi$ and $J/\psi$ mesons under their
respective collision
conditions~\cite{Abdallah_Nature2023,Acharya_PRL2023}. The
measured patterns motivate a comparison of the two mesons within a
common theoretical framework~\cite{Sheng_PRD2024}.

For a vector meson, the spin state is encoded in the $3\times3$
spin density matrix $\rho_{\lambda\lambda'}$, with
$\lambda,\lambda'=0,\pm1$ labeling spin projections along a chosen
quantization axis. The alignment is characterized by $\rho_{00}$,
the probability for zero spin projection. Its deviation from $1/3$
measures the departure from the unaligned value for that axis.
Spin alignment is extracted from decay angular distributions.
Relevant channels include strong $p$-wave decays such as $\phi\to
K^+K^-$~\cite{Abdallah_Nature2023,Acharya_PRL2020,Liang_PLB2005,Yang_PRC2018}
and dilepton decays such as
$J/\psi\to\mu^+\mu^-$~\cite{Acharya_PRL2023,Schilling_NPB1970,Faccioli_EPJC2010,Acharya_PLB2021}.
For the latter, the polar angle distribution is conventionally
parametrized by $\lambda_\theta$~\cite{Faccioli_EPJC2010}:
\begin{equation}
    W(\theta)\propto\frac{1}{3+\lambda_{\theta}}
    \left(1+\lambda_{\theta}\cos^{2}\theta\right),
\end{equation}
where $\theta$ is the angle between the daughter momentum and the
quantization axis in the parent rest frame. Neglecting the lepton
mass, the relation to $\rho_{00}$ for dilepton decays
is~\cite{Sheng_PRD2024,Zhao_JHEP2024}
\begin{equation}
    \lambda_{\theta}=\frac{1-3\rho_{00}}{1+\rho_{00}}.
\end{equation}
The limiting cases $\lambda_{\theta}=-1$ ($\rho_{00}=1$) and
$\lambda_{\theta}=1$ ($\rho_{00}=0$) correspond to purely
longitudinal and transverse polarization, respectively.

Vector meson spin alignment in rotating media has also been
studied in the NJL model and its Polyakov-loop
extension~\cite{Wei_CPC2023,Sun_PRD2024}. Holographic methods have
been used to study vector meson spin alignment through in-medium
current spectral functions~\cite{Ahmed_PRD2025,Ferreira_PRD2025}.
Zhao et al.~\cite{Zhao_JHEP2024} studied the $J/\psi$ spin
parameters in a thermal magnetized background. Sheng et
al.~\cite{Sheng_PRD2024} developed a framework relating
spin-resolved dilepton production rates to vector current spectral
functions. In their zero-density soft-wall calculation at
$T=0.15~\mathrm{GeV}$, the $J/\psi$ has a pronounced spectral
resonance whereas the $\phi$ does not. They obtained positive and
negative helicity-frame deviations from $1/3$ for the $J/\psi$ and
$\phi$, respectively.

The extension of this framework to finite density requires care,
and not every implementation gives the same answer. Chen et
al.~\cite{Chen_arXiv2026} recently studied spin alignment of
$K^{*}$, $\phi$, $D^{*}$, $D_s^{*}$, and $J/\psi$ in a unified
four-flavor soft-wall model with an anisotropic
Einstein--Maxwell--dilaton (EMD) background. Their background
includes finite temperature, baryon chemical potential, and
angular velocity. They find that heavy charm vector mesons exhibit
$\rho_{00}>1/3$ at low transverse momentum, while light strange
vector mesons show $\rho_{00}<1/3$, and they trace this flavor
separation to the location of the vacuum mass shell relative to
the thermally shifted spectral peaks. For the question of density
dependence, their conclusion is that spin alignment is insensitive
to the baryon chemical potential and only weakly affected by
angular velocity.

The present work approaches finite density from a different
parametrization. We use a charged black-hole background in which
the chemical potential is a quark chemical potential, normalized
through the horizon condition $\mu=q\zeta_h^2$, and the neutral
vector probe couples to the background only through the metric.
This is not a more fundamental choice than the baryon chemical
potential used in Ref.~\cite{Chen_arXiv2026}. It is a different
model input, and it leads to a quantitatively different density
response. In the range studied here, the spectral weights and
$\rho_{00}$ show visible changes with $\mu$, whereas the baryon
chemical potential implementation of Ref.~\cite{Chen_arXiv2026}
reports only a weak sensitivity. The two results are therefore
complementary rather than contradictory. They delimit how much of
the finite-density response is controlled by the choice of
background and chemical potential normalization.

The remainder of the calculation follows the soft-wall
construction of Ref.~\cite{Sheng_PRD2024}. The zero-density limit
reproduces the earlier spectral functions and spin alignment, and
the finite-density results are presented as a controlled
deformation of that benchmark. We give the background, the
equations of motion, the boundary kernels, and the spin projection
in full so that the calculation can be repeated with other dilaton
profiles or background charges. The qualitative features are
separated from the numerical shifts in the discussion.

The rest of this paper is structured as follows. Section~II sets
up the background geometry and the holographic model for vector
mesons and derives the equations of motion. Section~III relates
the dimuon production rate to spin alignment. Section~IV presents
numerical results, and Section~V gives a summary and outlook.

\section{Model setup}

The model and the calculational framework follow the soft-wall
construction of Ref.~\cite{Sheng_PRD2024}, generalized to a
finite-density plasma. This is a phenomenological bottom-up model.
It does not correspond to a known field theory. We state the full
set of equations so that the finite-density extension can be
checked independently.

\subsection{Finite-density plasma}

We begin with the black-hole background metric describing a system
at finite
density~\cite{Colangelo_PRD2011,Colangelo_EPJC2012,Colangelo_JHEP2012}:
\begin{equation}
    ds^2=\frac{R^2}{\zeta^2}
    (
    -f(\zeta)dt^2+d\vec{x}\cdot d\vec{x}
    +\frac{d\zeta^2}{f(\zeta)}
    ), \label{ds2}
\end{equation}
with
\begin{equation}
    f(\zeta)=1-\frac{\zeta^4}{\zeta_h^4}
    -q^2\zeta_h^2\zeta^4
    +q^2\zeta^6.
\end{equation}
The parameter $q$ is proportional to the black-hole charge and
controls the finite-density deformation of the background
geometry. The horizon position $\zeta_h$ obeys $f(\zeta_h)=0$. We
define the temperature, chemical potential, and current
correlators at the asymptotic AdS boundary $\zeta=0$. Regularity
of the Euclidean geometry at the horizon fixes the time period to
$\Delta t=4\pi/|f'(\zeta_h)|$, giving
\begin{equation}
    T=\frac{|f'(\zeta_h)|}{4\pi}
    =\frac{1}{\pi\zeta_h}-\frac{q^2\zeta_h^5}{2\pi}.
\end{equation}
We use the nonextremal branch $q^2\zeta_h^6<2$, with $T>0$. A UV
cutoff is used only as a numerical regulator approaching this
boundary.

The parameter $q$ is proportional to the black-hole charge and is
associated with the density of the medium. On the gauge-theory
side, it is related to the quark chemical potential $\mu$. The
chemical potential $\mu$ acts as a source for the quark-number
density operator $\bar{\psi}\gamma^{0}\psi$ and couples to the
quark-number density in the Lagrangian. In the holographic
description, this role is played by the temporal component $V_0$
of a bulk vector field. We consider a configuration in which only
the temporal component is nonzero, $V_0=U_0(\zeta)$ and
$V_\zeta=V_i=0$.

More generally, the temporal component may be written as
$U_0(\zeta)=\mu-\kappa q\zeta^2$, where the dimensionless
parameter $\kappa$ fixes the normalization relating the geometric
charge to the quark chemical potential. The boundary condition
$U_0(0)=\mu$ and the horizon condition $U_0(\zeta_h)=0$ give
$\mu=\kappa q\zeta_h^2$. Following Ref.~\cite{Colangelo_JHEP2012},
we set $\kappa=1$ and use
\begin{equation}
    \mu=q\zeta_h^2.
    \label{eq:chemical_potential}
\end{equation}
This choice fixes the chemical potential scale of the model.
$\kappa$ is not independently calibrated to QCD observables here.

Thus, once $\zeta_h$ and $q$ are specified, both the temperature
and the chemical potential are fixed, and their effects are
encoded in the metric of Eq.~\eqref{ds2}. Holographic approaches
have been extensively employed to study finite-temperature effects
and various aspects of heavy-flavor physics
\cite{Toniato_PRD2025,Branz_PRD2010,Gutsche_PRD2013,Fadafan_EPJC2011,Fadafan_NPB2012,Fadafan_EPJC2014,Afonin_PLB2013,Mamani_JHEP2014,Hashimoto_JHEP2015,Liu_PRD2017,Liu_PLB2017,BallonBayona_PRD2017,Dudal_PRD2018,Gutsche_PRD2019,Bohra_PLB2020,Zhang_PLB2019,Dudal_PRD2017}.

\subsection{Holographic description of vector mesons}

With the background geometry in place, we introduce vector mesons
as probe fields propagating in this background to investigate the
medium-induced modifications of their properties. In the probe
approximation, the backreaction of the vector mesons on the
geometry is neglected. We use the soft-wall vector action of
Ref.~\cite{Karch_PRD2006}:
\begin{equation}
    S_M=-\frac{1}{4g_5^2}\int d^4x\,d\zeta\,
    e^{-\phi}\sqrt{-g}\,
    F_{MN}F^{MN}, \label{Sm}
\end{equation}
where $F_{MN}=\nabla_M A_N-\nabla_N A_M=\partial_M A_N-\partial_N
A_M$ is the field strength tensor, $\nabla_M$ is the covariant
derivative, and $g_5$ is the five-dimensional gauge coupling. This
action and the holographic prescription below are the same as
those used in Ref.~\cite{Sheng_PRD2024}. The finite-density
background is the new element in the present work. To obtain the
correlation functions, we solve the equations of motion for the
five-dimensional vector field $A_\mu$. Variation of the action in
Eq.~\eqref{Sm} yields
\begin{equation}
    \partial_M\left(\frac{e^{-\phi}\sqrt{-g}}{4g_5^2}F^{MN}\right)=0. \label{eom}
\end{equation}
We adopt the radial gauge $A_\zeta=0$ and perform the Fourier
transformation
\begin{equation}
    A_\mu(x,\zeta)=\int\frac{d^4p}{(2\pi)^4}
    e^{-i\omega t+i\boldsymbol p\cdot\boldsymbol x}
    A_\mu(p,\zeta).
\end{equation}
Here $p_\mu=(-\omega,\boldsymbol p)$, and the boundary metric has
signature $\eta_{\mu\nu}=\mathrm{diag}(-1,1,1,1)$. The
gauge-invariant electric fields are
\begin{equation}
    E_i(p,\zeta)=-p_tA_i(p,\zeta)+p_iA_t(p,\zeta),
    \qquad i=x,y,z,\quad p_t=-\omega.
\end{equation}
To define the correlator kernel, introduce a UV cutoff
$\zeta=\epsilon$ and write
\[
E_i(p,\zeta)=\mathcal V_i{}^j(p,\zeta;\epsilon)E_j^\epsilon(p),
\qquad \mathcal V_i{}^j(p,\epsilon;\epsilon)=\delta_i{}^j.
\]
Each column of $\mathcal V$ satisfies the incoming-wave condition
at the horizon. The cutoff sources obey $E_i^\epsilon=C_i^\mu
A_\mu^\epsilon$, where $A_\mu^\epsilon(p)=A_\mu(p,\epsilon)$ and
$C_i^\mu=-p_t\delta_i^\mu+p_i\delta_t^\mu$. Define
\[
Q=\frac{e^{-\phi}\sqrt{-g}}{4g_5^2},\qquad
b^i=g^{i\alpha}p_\alpha,\qquad s=g^{\alpha\beta}p_\alpha p_\beta,
\]
where the metric components are those of the inverse bulk metric.
The unsymmetrized UV boundary kernels are
\[
\begin{aligned}
H_\epsilon^{ij}(p) &=\left.\frac{2Qg^{\zeta\zeta}}{p_t^2}
\left(g^{ik}-\frac{b^i b^k}{s}\right)
\partial_\zeta\mathcal V_k{}^j(p,\zeta;\epsilon)
\right|_{\zeta=\epsilon},\\
\mathcal F_\epsilon^{\mu\nu}(p) &=-C_i^\mu
H_\epsilon^{ij}(p)C_j^\nu.
\end{aligned}
\]
In the normalization of Ref.~\cite{Zhao_JHEP2024}, the retarded
correlator is obtained from the incoming-wave Son--Starinets
prescription~\cite{Son_JHEP2002} as
\begin{equation}
    D^{\mu\nu}(p)=2\lim_{\epsilon\to0}
    \mathcal F_\epsilon^{\mu\nu}(p).
    \label{eq:retarded_correlator}
\end{equation}
Only the UV kernel is retained in this prescription. The boundary
limit is understood after subtraction of real local contact terms,
which do not affect the spectral tensor~\cite{Son_JHEP2002}. The
kernel $\mathcal F_\epsilon^{\mu\nu}$ is distinct from the field
strength $F_{MN}$. The spectral tensor is
\begin{equation}
    \varrho^{\mu\nu}(p)=-\operatorname{Im}D^{\mu\nu}(p).
    \label{eq:spectral_tensor}
\end{equation}

\subsection{Equations of motion}

To evaluate the correlation functions numerically, we need to
solve Eq.~\eqref{eom}~\cite{Zhao_JHEP2024}. The component form of
Eq.~\eqref{eom} reads
\begin{equation}
    \begin{aligned}
        A_t''
        +\left(-\frac{1}{\zeta}-\phi'\right)A_t'
        -\frac{1}{f}
        [
        p_z(p_zA_t+\omega A_z)
        +(p_x^2+p_y^2)A_t
        +\omega(p_xA_x+p_yA_y)
        ]
        &=0,
        \\
        A_x''
        +\left(-\frac{1}{\zeta}+\frac{f'}{f}-\phi'\right)A_x'
        +\frac{\omega(p_xA_t+\omega A_x)}{f^2}
        +\frac{p_y(p_xA_y-p_yA_x)+p_z(p_xA_z-p_zA_x)}{f}
        &=0,
        \\
        A_y''
        +\left(-\frac{1}{\zeta}+\frac{f'}{f}-\phi'\right)A_y'
        +\frac{\omega(p_yA_t+\omega A_y)}{f^2}
        +\frac{p_x(p_yA_x-p_xA_y)+p_z(p_yA_z-p_zA_y)}{f}
        &=0,
        \\
        A_z''
        +\left(-\frac{1}{\zeta}+\frac{f'}{f}-\phi'\right)A_z'
        +\frac{p_z(p_xA_x+p_yA_y)-(p_x^2+p_y^2)A_z}{f}
        +\frac{\omega(p_zA_t+\omega A_z)}{f^2}
        &=0,
        \\
        \omega A_t'
        +f(p_xA_x'+p_yA_y'+p_zA_z')
        &=0.
    \end{aligned}
    \label{EOM_C}
\end{equation}
Here a prime denotes differentiation with respect to $\zeta$.
These five equations describe the coupling between the gauge-field
fluctuations and the finite-density background. The corresponding
electric-field components, $E_j=\omega A_j+p_jA_t$, obey
\begin{equation}
    \begin{aligned}
        E_x''
        +\left(-\frac{1}{\zeta}+\frac{f'}{f}-\phi'
        +p_x^2\mathcal{B}\right)E_x'
        +p_xp_y\mathcal{B}E_y'
        +p_xp_z\mathcal{B}E_z'
        +\mathcal{A}E_x
        &=0,
        \\
        E_y''
        +\left(-\frac{1}{\zeta}+\frac{f'}{f}-\phi'
        +p_y^2\mathcal{B}\right)E_y'
        +p_xp_y\mathcal{B}E_x'
        +p_yp_z\mathcal{B}E_z'
        +\mathcal{A}E_y
        &=0,
        \\
        E_z''
        +\left(-\frac{1}{\zeta}+\frac{f'}{f}-\phi'
        +p_z^2\mathcal{B}\right)E_z'
        +p_z\mathcal{B}(p_xE_x'+p_yE_y')
        +\mathcal{A}E_z
        &=0.
    \end{aligned}
\end{equation}
where
\begin{equation}
\begin{aligned}
    \mathcal{A}(\zeta)
    &=\frac{\omega^2}{f^2}
    -\frac{p_x^2+p_y^2+p_z^2}{f},
    \\
    \mathcal{B}(\zeta)
    &=\frac{f'}{\omega^2-f(p_x^2+p_y^2+p_z^2)}.
\end{aligned}
\end{equation}
Near the horizon, we impose an ingoing-wave condition. Thus
$E(\zeta)$ can be written as
\begin{equation}
    E(\zeta)=e^{-i\omega r_*}\psi(\zeta),\label{E_zeta}
\end{equation}
where $r_*$ is the tortoise coordinate defined by
$\partial_{r_*}=-f(\zeta)\partial_\zeta$. After a suitable
transformation, the system can be rewritten
as~\cite{Sheng_PRD2024,Zhao_JHEP2024}
\begin{equation}
    \left(
    \widetilde{I}\,\partial_\zeta^2
    +\widetilde{K}\,\partial_\zeta
    +\widetilde{N}
    \right)\psi(\zeta)=0.
\end{equation}
Here $\widetilde{I}$, $\widetilde{K}$, and $\widetilde{N}$ are
$3\times3$ matrices, and $\psi(\zeta)$ is a three-component
vector. Near the horizon, we expand $\psi(\zeta)$ as
\begin{equation}
    \psi(\zeta)=\psi_0+\sum_{n=1}^{\infty}a_n
    \left(\frac{\zeta}{\zeta_h}-1\right)^n. \label{taylor}
\end{equation}
The incoming initial conditions can be constructed from this
near-horizon expansion as in
Refs.~\cite{Sheng_PRD2024,Zhao_JHEP2024}. In the present isotropic
background, we solve the longitudinal and transverse modes
separately, using the leading incoming-wave behavior to initialize
the integration just outside the horizon.

\section{Dilepton production through vector meson decay}

The derivation in this section follows Ref.~\cite{Sheng_PRD2024}.
We recall it here to make the finite-density generalization
self-contained.

We consider dilepton production via the decay of a vector meson,
e.g., $J/\psi\to l+\bar l$ or $\phi\to l+\bar l$. For a transition
from an initial state $i$ to a final state $f$ accompanied by a
lepton pair $l\bar l$, the $S$-matrix element
is~\cite{Gale_NPB1991}
\begin{equation}
    S_{fi}=\int d^4x\,d^4y\,\langle f,l\bar l|J^\mu(y)G_{\mu\nu}^R(x-y)J_l^\nu(x)|i\rangle.
    \label{Sfi}
\end{equation}

Equation~\eqref{Sfi} describes a two-stage process. The vector
meson is produced from the initial state at $y$, propagates, and
then decays into a lepton pair at $x$. Here $G_{\mu\nu}^R$ is the
retarded propagator of the vector meson, satisfying
$G_{\mu\nu}^R(x-y)=0$ for $x^0-y^0<0$ to ensure causality.

In Eq.~\eqref{Sfi}, $J^\nu$ is the current coupled to the vector
meson. For the QGP, $J^\mu$ is the quark current. For a hadron
gas, it is the corresponding hadronic current. The leptonic
current $J_l^\mu$ is
\begin{equation}
    J_l^\nu(x)=g_{Ml\bar l}\bar\psi_l(x)\Gamma^\nu\psi_l(x),
    \label{lepton-current}
\end{equation}
where $\psi_l$ is the lepton field, $g_{Ml\bar l}$ the coupling
strength, and $\Gamma^\nu$ the effective interaction vertex. In
general, $\Gamma^\nu$ may be a linear combination of Lorentz
vectors such as
$\overrightarrow{\partial}^{\,\nu}-\overleftarrow{\partial}^{\,\nu}$,
$\gamma^\nu$, and
$\sigma^{\nu\alpha}(\overrightarrow{\partial}_\alpha-\overleftarrow{\partial}_\alpha)$.
For simplicity, we keep only the vector structure and take
$\Gamma^\nu\simeq\gamma^\nu$.

The transition probability per unit spacetime volume is
$R_{fi}\equiv |S_{fi}|^2/(TV)$, where $TV$ is the spacetime
volume. After summing over final states and thermally averaging
over initial states, the differential dilepton production rate
$n(x,p)\equiv dN/(d^4x\,d^4p)$ as a function of the total
four-momentum $p^\mu=(\omega,\boldsymbol p)$
is~\cite{Sheng_PRD2024}
\begin{align}
    n(x,p)={}&-\frac{2g_{Ml\bar l}^2}{3(2\pi)^5}(1-\frac{2m_l^2}{p^2})\sqrt{1+\frac{4m_l^2}{p^2}}\,p^2n_B(x,\omega) \notag\\
    &\times P_{\mu\nu}(p)G_A^{\mu\alpha}(p)\varrho_{\alpha\beta}(x,p)G_R^{\beta\nu}(p).
    \label{dilepton-rate}
\end{align}
This rate follows Ref.~\cite{Sheng_PRD2024}, with the transverse
projector written consistently with the metric convention as
\[
P_{\mu\nu}(p)=\eta_{\mu\nu}-\frac{p_\mu p_\nu}{p^2}, \qquad p^\mu
P_{\mu\nu}=0,\qquad p^2=-\omega^2+\boldsymbol p^2=-M^2.
\]
We consider the physical dilepton region $\omega>0$, $M>0$, and
$M\ge2m_l$. Here $m_l$ is the lepton mass, $G_A^{\mu\nu}$ is the
advanced propagator, and $n_B(x,\omega)=1/(e^{\omega/T(x)}-1)$ is
the Bose--Einstein distribution at the local temperature $T(x)$.
The in-medium spectral tensor is
\begin{equation}
    \varrho_{\alpha\beta}(x,p)\equiv-\operatorname{Im}D_{\alpha\beta}(x,p),
    \label{spectral-function}
\end{equation}
with the retarded current-current correlator
\begin{equation}
    D^{\mu\nu}(x,p)\equiv-i\int d^4y\,\theta(y^0)
    \langle[J^\mu(y),J^\nu(0)]\rangle_{T(x),\mu}
    e^{-ip\cdot y}.
    \label{current-correlator}
\end{equation}
The notation $\langle O\rangle_{T(x),\mu}$ denotes the thermal
average at the given local temperature $T(x)$ and quark chemical
potential $\mu$, with $\mu$ held fixed in each calculation.

It is important to distinguish the roles of the quantities in
Eq.~\eqref{dilepton-rate}. The tensor $\varrho_{\alpha\beta}$
describes the spectral properties of the vector meson in the
thermal medium, whereas $G_{R/A}^{\mu\nu}$ are vacuum propagators
that describe its free propagation after freezeout. We neglect the
detailed kinematics of freezeout and adopt an instantaneous
freezeout approximation, in which the spectral function
$\varrho_{\alpha\beta}$ is sandwiched between the free
propagators, $G_A^{\mu\alpha}\varrho_{\alpha\beta}G_R^{\beta\nu}$.

If we replace the retarded and advanced propagators by the vacuum
photon propagator $\eta^{\mu\nu}/p^2$, Eq.~\eqref{dilepton-rate}
reduces to the standard dilepton rate from photon
decay~\cite{Gale_NPB1991,Burnier_JHEP2009,McLerran_PRD1985,Weldon_PRD1990}.

Current conservation ensures that the dilepton rate depends only
on the four-dimensional transverse part of the propagator. In the
positive-frequency resonance region, $\omega>0$, we use the
effective Breit--Wigner form
\begin{equation}
    G_{R/A}^{\mu\nu}(p)=
    -\frac{P^{\mu\nu}(p)}
    {p^2+m_V^2\mp im_V\Gamma},
    \qquad M^2=-p^2.
    \label{vector-propagator}
\end{equation}
The upper (lower) sign corresponds to the retarded (advanced)
propagator, and $P^{\mu\nu}=\eta^{\mu\nu}+p^\mu p^\nu/M^2$. Here
$m_V$ is the vacuum meson mass and $\Gamma$ is the width parameter
of the effective propagator.

The spectral function can be decomposed in a basis of polarization
vectors as
\begin{equation}
    \varrho^{\mu\nu}(x,p)=\sum_{\lambda,\lambda'=0,\pm1}v^\mu(\lambda,p)v^{*\nu}(\lambda',p)\widetilde{\varrho}_{\lambda\lambda'}(x,p), \label{varrho_uv}
\end{equation}
with the polarization vectors
\begin{equation}
    v^\mu(\lambda,p)=(\frac{\boldsymbol p\cdot\boldsymbol\epsilon_\lambda}{M},\boldsymbol\epsilon_\lambda+\frac{\boldsymbol p\cdot\boldsymbol\epsilon_\lambda}{M(\omega+M)}\boldsymbol p),
    \label{polarizationvector}
\end{equation}
which satisfy the orthonormality relation
$\eta_{\mu\nu}v^\mu(\lambda,p)v^{*\nu}(\lambda',p)
=\delta_{\lambda\lambda'}$ and the completeness relation
\[
\sum_\lambda v^\mu(\lambda,p)v^{*\nu}(\lambda,p)
=\eta^{\mu\nu}-\frac{p^\mu p^\nu}{p^2} =\eta^{\mu\nu}+\frac{p^\mu
p^\nu}{M^2}.
\]
Here $M=\sqrt{\omega^2-\boldsymbol p^2}$ is the invariant mass.
The three-dimensional vectors $\boldsymbol\epsilon_\lambda$
specify the spin directions in the meson rest frame, with
$\boldsymbol\epsilon_0$ defining the quantization direction and
$\boldsymbol\epsilon_{\pm1}$ orthogonal to it.

If the spin quantization axis is chosen along $\boldsymbol p$, the
$\lambda=\pm1$ and $\lambda=0$ contributions correspond to the
transverse and longitudinal projections, respectively. The
dilepton rate then decomposes into spin-state
contributions~\cite{Sheng_PRD2024}:
\begin{align}
    n_\lambda(x,p)={}&-\frac{2g_{Ml\bar l}^2}{3(2\pi)^5}(1-\frac{2m_l^2}{p^2})\sqrt{1+\frac{4m_l^2}{p^2}} \notag\\
    &\times\frac{p^2n_B(x,\omega)\widetilde{\varrho}_{\lambda\lambda}(x,p)}{(p^2+m_V^2)^2+m_V^2\Gamma^2}.
    \label{spin-resolved-rate}
\end{align}
In the zero-density limit, Eq.~\eqref{spin-resolved-rate} and the
resulting $\rho_{00}$ reduce to the results first obtained in
Ref.~\cite{Sheng_PRD2024}. The total rate is
$n(x,p)=\sum_{\lambda=0,\pm1}n_\lambda(x,p)$. For a vector meson
resonance, the spin alignment parameter is the fraction of the
dilepton yield in the zero projection state~\cite{Sheng_PRD2024}:
\begin{equation}
    \rho_{00}(x,\boldsymbol p)\equiv\frac{\int d\omega\,n_0(x,p)}{\sum_{\lambda=0,\pm1}\int d\omega\,n_\lambda(x,p)}.
    \label{spin-alignment}
\end{equation}
The energy integration in Eq.~\eqref{spin-alignment} is at fixed
three-momentum $\boldsymbol p$. In terms of the invariant mass,
$d\omega=(M/\omega)dM$, with $\omega=\sqrt{M^2+|\boldsymbol
p|^2}$. Since each $n_\lambda$ is nonnegative,
$0\leq\rho_{00}\leq1$. For a narrow resonance, $\Gamma\ll m_V$,
the denominator in Eq.~\eqref{spin-resolved-rate} peaks sharply at
$-p^2=m_V^2$. When the remaining factors vary slowly across this
region, $\rho_{00}$ is approximately given by the ratio
$\widetilde{\varrho}_{00}/\sum_{\lambda}\widetilde{\varrho}_{\lambda\lambda}$
at $\omega=\sqrt{m_V^2+|\boldsymbol p|^2}$.

\section{Numerical results}
Using the narrow-width approximation stated in
Ref.~\cite{Sheng_PRD2024}, we evaluate the numerical spin
alignment parameters from spectral ratios at $M=m_V$, with all
kinematic factors evaluated at the same mass. We set the AdS
radius $R=1$ in the numerical calculation.

\subsection{$J/\psi$ meson}

We begin with the dimuon decay channel of the $J/\psi$. We use the
vacuum mass $m_{J/\psi}=3.097~\mathrm{GeV}$~\cite{Navas_PRD2024}.
The dilaton profile in Eq.~\eqref{Sm} is chosen as
$\phi(\zeta)=c_{J/\psi}\zeta^2$, with
$c_{J/\psi}=m_{J/\psi}^2/4\simeq2.40~\mathrm{GeV}^2$ as in
Ref.~\cite{Sheng_PRD2024}, using the quadratic soft-wall
spectrum~\cite{Karch_PRD2006}.

\begin{figure}[htbp]
    \centering
    \includegraphics[width=8cm]{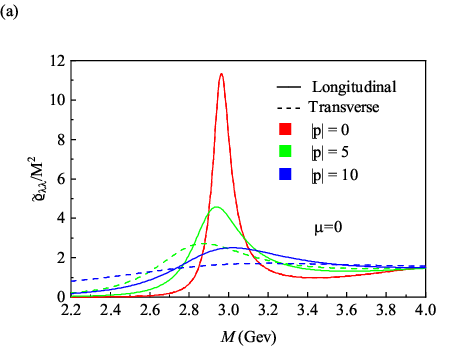}
    \includegraphics[width=8cm]{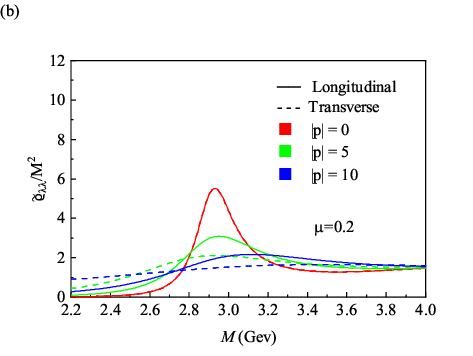}
    \includegraphics[width=8cm]{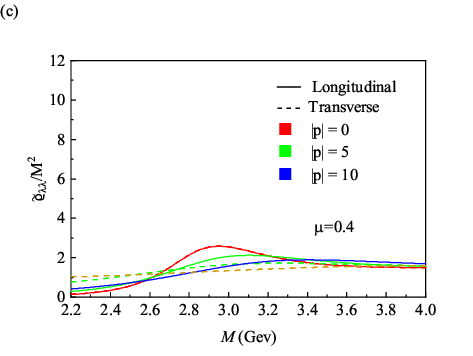}
    \includegraphics[width=8cm]{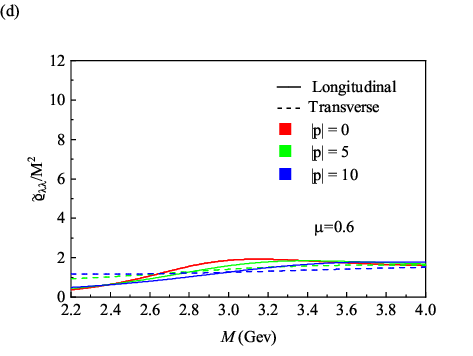}
    \caption{
        Spectral functions $\widetilde{\varrho}_{\lambda\lambda}$ of $J/\psi$ mesons as functions of the invariant mass
        $M\equiv\sqrt{\omega^{2}-\mathbf{p}^{2}}$
        at $T=0.15~\mathrm{GeV}$.
        Panels (a)--(d) correspond to chemical potentials
        $\mu=0$, $0.2$, $0.4$, and $0.6~\mathrm{GeV}$, respectively.
        In each panel, the red, green, and blue curves represent
        $|\mathbf{p}|=0$, $5$, and $10~\mathrm{GeV}$, respectively.
        Solid and dashed curves denote the longitudinal and transverse spectral functions, respectively.
        The displayed spectra are $g_5^2\widetilde{\varrho}_{\lambda\lambda}/M^2$. The common factor $g_5^2$ is suppressed in the vertical axis labels.
    }
\end{figure}

For a $J/\psi$ propagating through the thermal medium, the natural
spin quantization axis is the direction of its three-momentum in
the local rest frame of the medium,
$\boldsymbol{\epsilon}_0=\boldsymbol{e}_0^h=\mathbf{p}/|\mathbf{p}|$.
This defines the helicity frame (HF). In the HF, the spatial
component of the longitudinal polarization vector $\lambda=0$ is
aligned with $\mathbf{p}$. The finite motion of the meson relative
to the medium breaks the degeneracy between the longitudinal and
the two transverse modes $\lambda=\pm1$, leading to different
dilepton rates $n_\lambda$ and thus nonzero spin alignment. Fig.~1
shows the spectral functions at $T=0.15~\mathrm{GeV}$ for several
chemical potentials and momenta, plotted against the invariant
mass $M$.

For $\mu=0$, the spectral functions and the resulting $\rho_{00}$
shown in Figs.~1 and 2 reproduce the zero-density results of
Ref.~\cite{Sheng_PRD2024}, where the $J/\psi$ spin alignment was
first computed in this holographic framework. Our finite-density
results approach this limit smoothly as $\mu\to0$.

We first examine the momentum dependence. At $|\mathbf{p}|=0$,
rotational invariance enforces the degeneracy of the three spin
states, as seen by the red curve. At finite momentum, the
longitudinal spectral function (solid curves) separates from the
transverse one (dashed curves). In the low invariant-mass region,
the transverse modes dominate, while the longitudinal contribution
grows with increasing $M$. The resonance peaks decrease in height
as $|\mathbf p|$ increases. As $\mu$ increases, the peaks become
lower and broader, while their positions in $M$ remain
approximately unchanged. These statements describe the calculated
spectral response.

\begin{figure}[htbp]
    \centering
    \includegraphics[width=8cm]{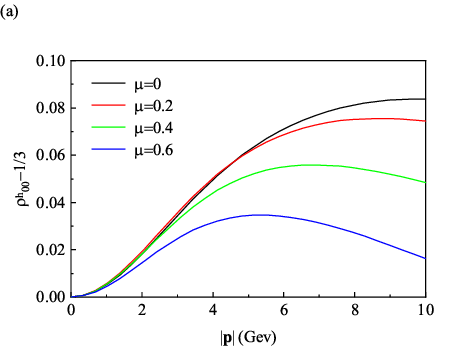}
    \includegraphics[width=8cm]{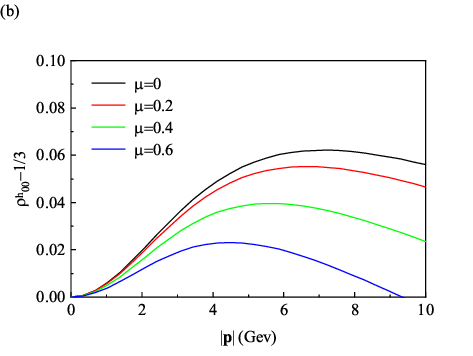}
    \caption{
    Spin alignment of the $J/\psi$ meson in the helicity frame (HF) as a function of momentum $|\mathbf{p}|$. (a) $T=0.15~\mathrm{GeV}$; (b) $T=0.2~\mathrm{GeV}$.  In each panel, the black, red, green, and blue curves correspond to chemical potentials $\mu=0$, $0.2$, $0.4$, and $0.6~\mathrm{GeV}$, respectively.
    }
\end{figure}

The HF spin alignment is evaluated from the on-shell spectral
ratio specified above. Fig.~2 shows its momentum dependence at
different temperatures and chemical potentials. At $\mu=0$,
$\rho_{00}>1/3$ for nonzero momenta shown and $\rho_{00}=1/3$ at
zero momentum. At $T=0.15~\mathrm{GeV}$, changing $\mu$ from $0$
to $0.2~\mathrm{GeV}$ raises the low-momentum curve but lowers it
at higher momenta. Further increases to $0.4$ and
$0.6~\mathrm{GeV}$ lower the curves over the nonzero momenta
shown. At $T=0.2~\mathrm{GeV}$, increasing $\mu$ lowers
$\rho_{00}$, and the $\mu=0.6~\mathrm{GeV}$ curve crosses below
$1/3$ at large momentum. The corresponding $\lambda_\theta$
changes from negative to positive.

To investigate global spin alignment in heavy-ion collisions, we
choose the quantization axis along the $y$ direction,
$\boldsymbol\epsilon_0=\boldsymbol e_0^y=(0,1,0)$, identified with
the direction of the global orbital angular momentum. In the
isotropic background, the spectral matrix is diagonal in the
helicity basis and the two transverse components are degenerate.
Define the unnormalized trace and the rotation factor as
\[
S(x,p)=\sum_\lambda
\widetilde{\varrho}_{\lambda\lambda}^{\,h}(x,p), \qquad
u=\boldsymbol e_0^y\cdot\boldsymbol e_0^h.
\]
Equation~\eqref{varrho_uv} then gives
\begin{align}
    \widetilde{\varrho}_{00}^{\,y}(x,p)
    &=
    \frac12\left\{
    S(x,p)-\widetilde{\varrho}_{00}^{\,h}(x,p)
    +u^2\left[3\widetilde{\varrho}_{00}^{\,h}(x,p)-S(x,p)\right]
    \right\},
    \nonumber\\
    \sum_\lambda\widetilde{\varrho}_{\lambda\lambda}^{\,y}(x,p)
    &=S(x,p).
\end{align}
Here $\widetilde{\varrho}_{\lambda\lambda'}$ denotes unnormalized
spectral weights, whereas $\rho_{00}$ is the normalized spin
alignment parameter. At fixed nonzero three-momentum, combining
these relations with Eqs.~\eqref{spin-resolved-rate}
and~\eqref{spin-alignment} yields
\begin{equation}
    \rho_{00}^{\,y}(x,\boldsymbol p)
    =
    \frac13+
    \frac{3\rho_{00}^{\,h}(x,\boldsymbol p)-1}
    {3|\boldsymbol p|^2}
    \left(p_y^2-\frac{p_x^2+p_z^2}{2}\right).
\end{equation}
At zero momentum, the isotropic continuous limit is
$\rho_{00}=1/3$. This spin basis projection follows Eqs.~(31) and
(32) of Ref.~\cite{Sheng_PRD2024}. At fixed $p_T$ and $Y$, using
the fixed vacuum-mass prescription $M=m_V$ specified above, define
\[
A=\rho_{00}^{h}-\frac13,\qquad
F=\frac{p_T^2}{p_T^2+(m_V^2+p_T^2)\sinh^2Y}.
\]
Substitution of the momentum parametrization into the projection
relation gives
\[
\rho_{00}^{y}-\frac13=c_1+c_2\cos(2\varphi),\qquad
c_1=A\left(\frac{3F}{4}-\frac12\right),\quad c_2=-\frac{3AF}{4}.
\]
Thus the azimuthal structure follows from the basis rotation,
while the coefficients depend on the calculated longitudinal and
transverse spectral weights.

\begin{figure}[htbp]
    \centering
    \includegraphics[width=8cm]{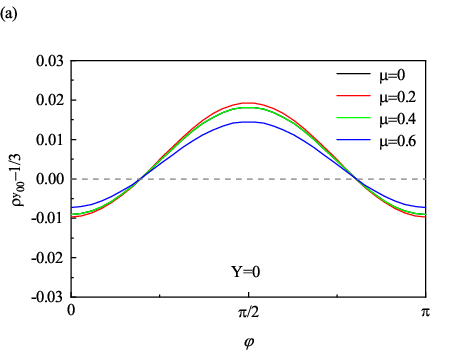}
    \includegraphics[width=8cm]{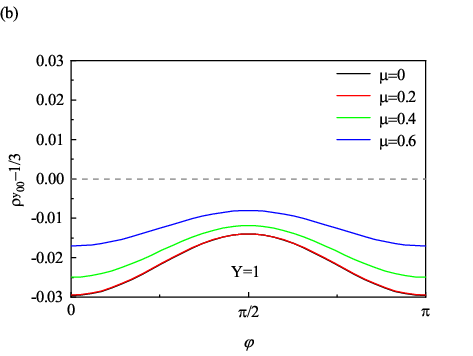}
    \caption{
    The global spin alignment of $J/\psi$ mesons with transverse momentum
    $p_T=2~\mathrm{GeV}$ as a function of the meson azimuthal angle $\varphi$.
    The left and right panels correspond to $Y=0$ and $Y=1$, respectively. The
    black, red, green, and blue curves represent chemical potentials $\mu=0$,
    $0.2$, $0.4$, and $0.6~\mathrm{GeV}$, respectively.
    }
\end{figure}

In Fig.~3, we show the global-frame quantity $\rho_{00}^{\,y}$ as
a function of the meson azimuthal angle. We assume a static
thermal medium, fix $p_T=2~\mathrm{GeV}$, and obtain the
longitudinal momentum at $M=m_{J/\psi}$ from
$p_z=\sqrt{m_{J/\psi}^2+p_T^2}\sinh Y$. The helicity spectra and
the spin basis rotation are evaluated at this same momentum. The
temperature is $T=0.15~\mathrm{GeV}$, and the chemical potential
takes values $\mu=0,0.2,0.4,0.6~\mathrm{GeV}$. For mesons at
midrapidity $Y=0$, $\rho_{00}^{\,y}$ is symmetric about
$\varphi=\pi/2$. At $\varphi=0$ it is below $1/3$. As $\varphi$
increases, it rises and exceeds $1/3$, reaching a maximum near
$\varphi=\pi/2$. The curves at $\mu=0$, $0.2$, and
$0.4~\mathrm{GeV}$ lie close together. Relative to $\mu=0$, the
$\mu=0.2~\mathrm{GeV}$ curve has a slightly higher maximum and
more negative endpoints, while the $\mu=0.4~\mathrm{GeV}$ curve
nearly coincides with the zero-density result. At
$\mu=0.6~\mathrm{GeV}$, both the positive maximum and the
magnitude of the negative endpoint deviations are reduced.

At $Y=1$, $\rho_{00}^{\,y}$ remains below $1/3$ over the whole
azimuthal range. Increasing $\mu$ shifts the plotted curves
upward, with the $\mu=0.6~\mathrm{GeV}$ curve closest to $1/3$.
The change from $\mu=0$ to $0.2~\mathrm{GeV}$ is small, while the
shifts at $0.4$ and $0.6~\mathrm{GeV}$ are more pronounced.

\begin{figure}[htbp]
    \centering
    \includegraphics[width=8cm]{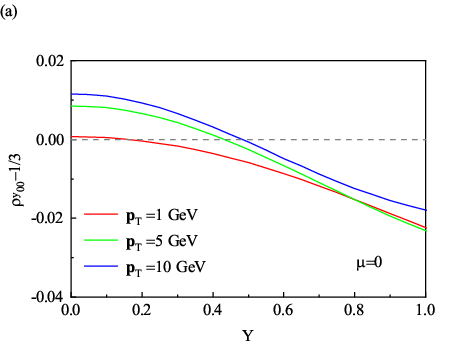}
    \includegraphics[width=8cm]{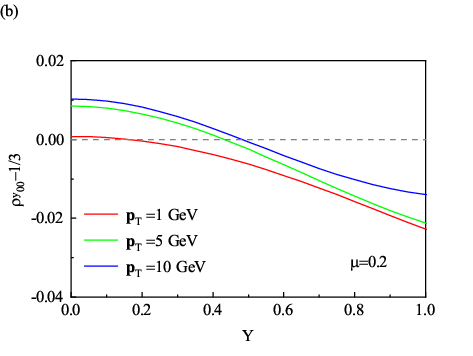}
    \includegraphics[width=8cm]{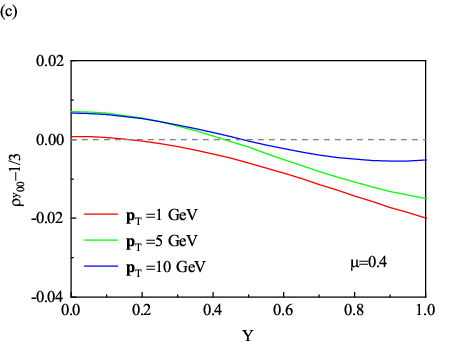}
    \includegraphics[width=8cm]{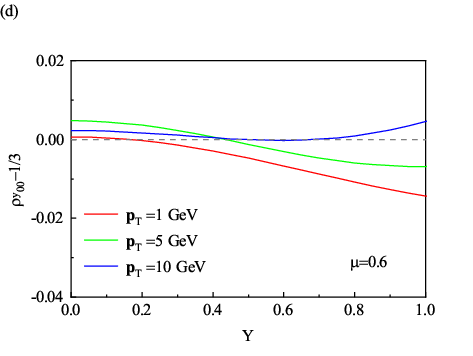}
    \caption{
        Global spin alignment of $J/\psi$ mesons as a function of the
        rapidity $Y$ for different chemical potentials. Panels (a)--(d)
        correspond to $\mu=0$, $0.2$, $0.4$, and $0.6~\mathrm{GeV}$,
        respectively. The red, green, and blue curves represent the meson
        transverse momenta $p_T=1$, $5$, and $10~\mathrm{GeV}$, respectively.
       }
\end{figure}

After averaging over the azimuthal angle $\varphi$, Fig.~4
displays the global spin alignment of the $J/\psi$ as a function
of rapidity for several momenta and chemical potentials. To
incorporate the anisotropic flow $v_2$ from the initial collision
geometry, we perform the average with the weight
$1+2v_2\cos(2\varphi)$ and set $v_2=0.15$~\cite{Sheng_PRD2024}.

At midrapidity, $\rho_{00}^{\,y}$ increases with $p_T$ for the
parameters shown, while the ordering of the momentum curves
changes at larger $Y$. For relatively small $\mu$, the alignment
is above $1/3$ at midrapidity and below $1/3$ at larger rapidity.
Some of the large $\mu$ curves change sign more than once as $Y$
increases. These results combine the calculated spectral imbalance
with the spin basis projection. A quantitative comparison with
ALICE measurements would require matching the collision conditions
and kinematic acceptance and specifying the medium evolution.

\subsection{$\phi$ meson}

We now apply the same framework to the $\phi$ meson. We again use
the quadratic dilaton $\phi(\zeta)=c_{\phi}\zeta^{2}$, with
$c_{\phi}=m_{\phi}^{2}/4\approx 0.26~\mathrm{GeV}^{2}$ using
$m_\phi=1.02~\mathrm{GeV}$~\cite{Sheng_PRD2024}. The spin
alignment is evaluated at $M=m_\phi$ with the same narrow-width
prescription.

Although the dimuon channel is not the dominant decay mode of the
$\phi$, we continue to consider this process and assume that the
spin alignment is either independent of the decay channel or only
weakly sensitive to it within the narrow-width
approximation~\cite{Sheng_PRD2024}. As noted below
Eq.~\eqref{spin-alignment}, at $M=m_\phi$, the spin alignment
parameter is approximately
\begin{equation}
    \rho_{00}\approx\frac{\widetilde{\varrho}_{00}}{\sum_{\lambda}\widetilde{\varrho}_{\lambda\lambda}},
\end{equation}
and the spectral function $\widetilde{\varrho}_{\lambda\lambda}$
itself does not depend on the decay process.

\begin{figure}[htbp]
    \centering
    \includegraphics[width=8cm]{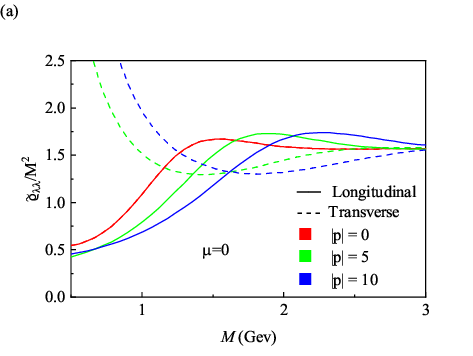}
    \includegraphics[width=8cm]{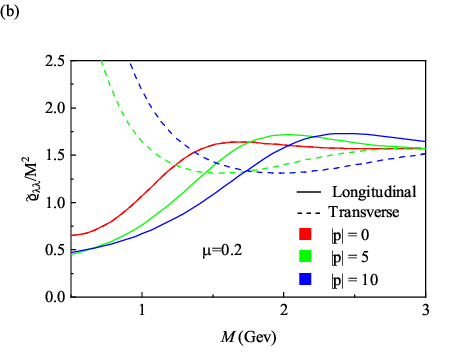}
    \includegraphics[width=8cm]{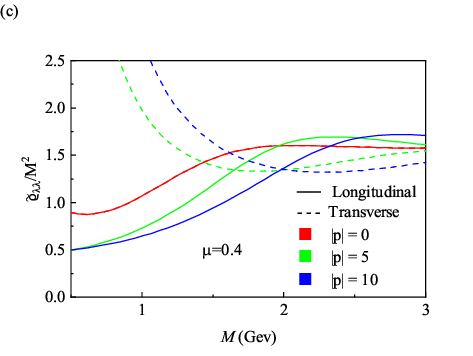}
    \includegraphics[width=8cm]{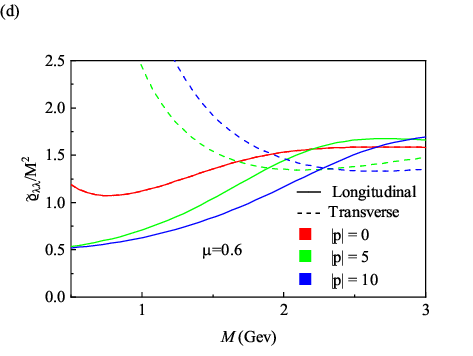}
    \caption{
        Spectral functions $\widetilde{\varrho}_{\lambda\lambda}$ of $\phi$ mesons as functions of the invariant mass
        $M\equiv\sqrt{\omega^{2}-\mathbf{p}^{2}}$
        at $T=0.15~\mathrm{GeV}$.
        Panels (a)--(d) correspond to chemical potentials
        $\mu=0$, $0.2$, $0.4$, and $0.6~\mathrm{GeV}$, respectively.
        In each panel, the red, green, and blue curves represent
        $|\mathbf{p}|=0$, $5$, and $10~\mathrm{GeV}$, respectively.
        Solid and dashed curves denote the longitudinal and transverse spectral functions, respectively.
        The displayed spectra are $g_5^2\widetilde{\varrho}_{\lambda\lambda}/M^2$. The common factor $g_5^2$ is suppressed in the vertical axis labels.
    }
\end{figure}

Fig.~5 presents the longitudinal and transverse spectral functions
of the $\phi$ meson as functions of $M$ at $T=0.15~\mathrm{GeV}$.
In contrast to the $J/\psi$ spectra in Fig.~1, no prominent $\phi$
resonance peak is observed. Following Ref.~\cite{Sheng_PRD2024},
we interpret these spectra as the relative spectral weight of
$s\bar{s}$ pairs with the corresponding quantum numbers in the
medium.

Following Ref.~\cite{Sheng_PRD2024}, we adopt an instantaneous
freezeout prescription in which the in-medium spectral tensor is
sandwiched between vacuum meson propagators, as in
Eq.~\eqref{dilepton-rate}. This prescription connects the
in-medium $s\bar{s}$ spectrum to the final meson decay channel
without resolving the dynamics of the freezeout stage. The
numerical spin alignment is evaluated at $M=m_\phi$ with the
narrow-width approximation specified above.

\begin{figure}[htbp]
    \centering
    \includegraphics[width=8cm]{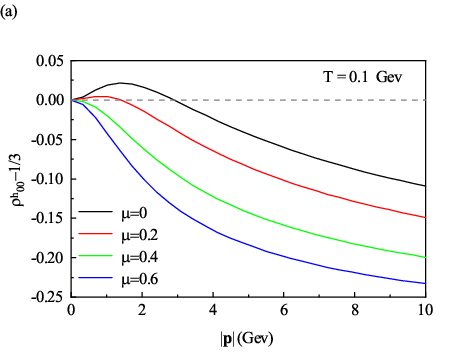}
    \includegraphics[width=8cm]{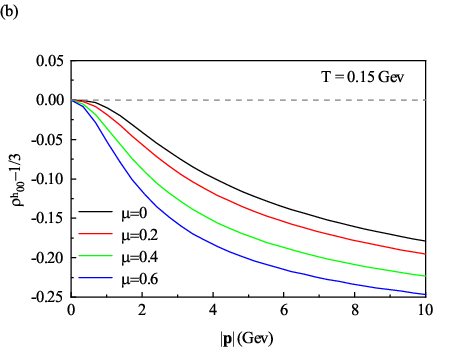}
    \caption{
        Spin alignment of the $\phi$ meson in the helicity frame as a function of momentum $|\mathbf{p}|$. (a) $T=0.1~\mathrm{GeV}$; (b) $T=0.15~\mathrm{GeV}$.  In each panel, the black, red, green, and blue curves correspond to chemical potentials $\mu=0$, $0.2$, $0.4$, and $0.6~\mathrm{GeV}$, respectively.
    }
\end{figure}

Fig.~6 shows the HF spin alignment parameter $\rho_{00}^{h}$ of
the $\phi$ meson as a function of momentum and chemical potential.
At $T=0.1~\mathrm{GeV}$ and for relatively small $\mu$,
$\rho_{00}^{h}$ has a nonmonotonic dependence on $|\mathbf{p}|$.
It first increases, reaches a maximum, and then decreases. At
$T=0.15~\mathrm{GeV}$, however, $\rho_{00}^{h}$ is below $1/3$ and
its negative deviation grows with increasing $|\mathbf{p}|$. As
$\mu$ increases, $\rho_{00}^{h}$ shifts to lower values. Both
temperature and chemical potential can modify the magnitude, sign,
and momentum dependence of the deviation, with qualitatively
similar trends.

At $\mu=0$, our $\phi$ meson spectral functions and
$\rho_{00}^{h}$ agree with the zero-density results first obtained
in Ref.~\cite{Sheng_PRD2024}. The finite-density curves presented
here therefore connect continuously to that benchmark.

\begin{figure}[htbp]
    \centering
    \includegraphics[width=8cm]{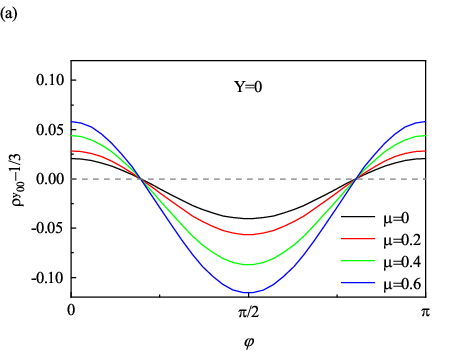}
    \includegraphics[width=8cm]{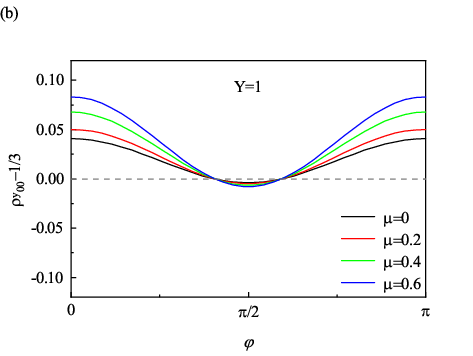}
    \caption{
    The global spin alignment of $\phi$ mesons with transverse momentum
    $p_T=2~\mathrm{GeV}$ as a function of the meson azimuthal angle $\varphi$.
    The left and right panels correspond to $Y=0$ and $Y=1$, respectively. The
    black, red, green, and blue curves represent chemical potentials $\mu=0$,
    $0.2$, $0.4$, and $0.6~\mathrm{GeV}$, respectively.
    }
\end{figure}

Fig.~7 displays the azimuthal-angle dependence of the global spin
alignment parameter $\rho_{00}^{\,y}-1/3$ for the $\phi$ meson at
$p_T=2~\mathrm{GeV}$ and $T=0.15~\mathrm{GeV}$. At $Y=0$,
$\rho_{00}^{\,y}-1/3$ is positive near $\varphi=0$ and
$\varphi=\pi$, decreases as $\varphi$ approaches $\pi/2$, and
reaches a negative minimum at $\pi/2$. As $\mu$ increases from $0$
to $0.6~\mathrm{GeV}$, the positive deviations near the endpoints
become larger, and the negative minimum becomes more pronounced.
At $Y=1$, the azimuthal dependence is similar, with larger values
near the endpoints and a negative minimum at $\pi/2$. Compared to
$Y=0$, the curves at $Y=1$ are shifted upward overall. Increasing
$\mu$ further enhances the azimuthal modulation, amplifying both
the positive deviations and the negative dip.

\begin{figure}[htbp]
    \centering
    \includegraphics[width=8cm]{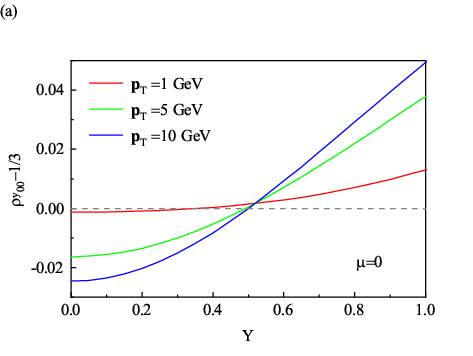}
    \includegraphics[width=8cm]{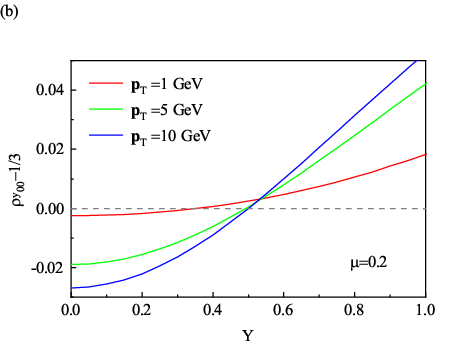}
    \includegraphics[width=8cm]{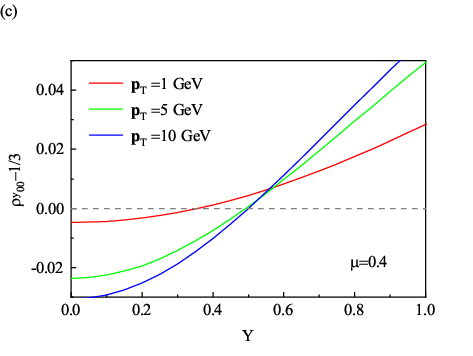}
    \includegraphics[width=8cm]{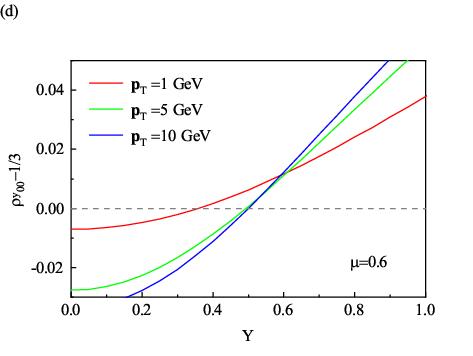}
    \caption{
    Global spin alignment of $\phi$ mesons as a function of the
    rapidity $Y$ for different chemical potentials. Panels (a)--(d)
    correspond to $\mu=0$, $0.2$, $0.4$, and $0.6~\mathrm{GeV}$,
    respectively. The red, green, and blue curves represent the meson
    transverse momenta $p_T=1$, $5$, and $10~\mathrm{GeV}$, respectively.
    }
\end{figure}

We then perform the azimuthal angle average with weight
$1+2v_2\cos(2\varphi)$, $v_2=0.15$. Fig.~8 shows the global spin
alignment $\rho_{00}^{\,y}$ of the $\phi$ meson as a function of
rapidity $Y$ and chemical potential $\mu$ for
$p_T=1,5,10~\mathrm{GeV}$.

In the parameter range shown, the calculated $\phi$ meson
alignment changes from below to above $1/3$ as rapidity increases.
Its deviation reaches order $10^{-2}$, which is comparable in
magnitude to the cited STAR measurement and coalescence model
result~\cite{Abdallah_Nature2023,Sheng_PRC2023}. This is an order
of magnitude comparison. A quantitative test requires matched
collision conditions and kinematic ranges.

\section{Summary and discussion}

In this work, we have extended the holographic study of
Ref.~\cite{Sheng_PRD2024} to flavorless vector mesons in a plasma
at finite density. For dilepton decays, the production rates in
different spin channels are governed by the imaginary part of the
retarded current-current correlator, i.e. the in-medium spectral
function. Within the gauge/gravity duality, the computation of
such correlators in a strongly coupled quantum field theory
reduces to solving a classical bulk field problem in AdS
spacetime. We derived the equations of motion for the bulk vector
field and reformulated them in terms of the electric field
components. With the infalling-wave boundary condition imposed at
the horizon, the retarded Green's function is extracted from the
boundary values of the fields and their radial derivatives.

We employed the soft-wall model to study the spin alignment of the
$J/\psi$ and $\phi$ mesons. In the isotropic background, the
motion of a meson relative to the thermal bath can produce
different longitudinal and transverse spectral weights. At $\mu=0$
and $T=0.15~\mathrm{GeV}$, the resulting $\rho_{00}^{h}>1/3$ for
the $J/\psi$ and $\rho_{00}^{h}<1/3$ for the $\phi$ reproduce the
qualitative patterns reported in Ref.~\cite{Sheng_PRD2024}. The
present extension concerns how these spectral weights change in
the charged background.

We further examined the influence of a finite quark chemical
potential on the spectral functions and spin alignment. At
$T=0.15~\mathrm{GeV}$, we computed the longitudinal and transverse
spectral functions for $\mu=0,0.2,0.4,0.6~\mathrm{GeV}$ and
momenta $|\mathbf{p}|=0,5,10~\mathrm{GeV}$. For the $J/\psi$, the
resonance peaks become progressively lower and broader as $\mu$
increases, indicating that a finite chemical potential enhances
dissociation and drives the $J/\psi$ toward melting. For the
$\phi$ meson, no pronounced resonance peak is observed at this
temperature over the chemical potentials shown.

We also studied global spin alignment along a fixed $y$ axis. In
this isotropic background, its $\cos(2\varphi)$ dependence follows
from the spin basis projection, with coefficients determined by
the spectral weights. The chemical potential changes these
coefficients, with the size and direction of the changes depending
on the meson species, momentum, and rapidity. These results
describe the model and kinematic range considered here. The
forward rapidity acceptance of the cited ALICE measurement is
outside that range.

At $\mu=0$ and $T=0.15~\mathrm{GeV}$, Fig.~1(a) shows a $J/\psi$
resonance, whereas Fig.~5(a) has no pronounced $\phi$ peak.
Following Ref.~\cite{Sheng_PRD2024}, the former can be interpreted
as the spectral weight of a quasistable meson, while the latter
characterizes $s\bar{s}$ pairs with the corresponding quantum
numbers in the medium. The finite-density spectra describe how
these structures change in the charged background.

We use the instantaneous freezeout prescription of
Ref.~\cite{Sheng_PRD2024}, which sandwiches the in-medium spectral
tensor between vacuum meson propagators. Within this prescription,
the vacuum propagators weight the spectrum near $M=m_V$, and the
numerical spin alignment uses the corresponding vacuum-mass-shell
spectral ratio.

The results of this work can be separated into robust qualitative
features and model-dependent numbers. The qualitative features are
the resonance peak of the $J/\psi$ and its absence for the $\phi$
at $T=0.15~\mathrm{GeV}$, the positive sign of $\rho_{00}^{h}-1/3$
for the $J/\psi$ and the negative sign for the $\phi$ at nonzero
momentum in the zero-density limit, and the visible lowering and
broadening of the $J/\psi$ peak with increasing quark chemical
potential in this charged background. These features follow from
the spectral weights and the spin basis projection, and they are
stable against the details of the dilaton profile within this
model class. The numerical values of $\rho_{00}$ and the detailed
$\mu$ dependence depend on the charged background, the quadratic
dilaton profile, the normalization $\kappa=1$, and the
instantaneous freezeout prescription.

The comparison with Ref.~\cite{Chen_arXiv2026} sharpens this
boundary. That work uses an anisotropic EMD background with baryon
chemical potential and angular velocity in a four-flavor soft-wall
model, and finds that spin alignment is insensitive to the baryon
chemical potential. The present model uses a quark chemical
potential through the metric of a charged black hole and finds
visible changes in the same range. The difference does not
indicate an inconsistency. It shows that the density response of
spin alignment is controlled by how the chemical potential enters
the background and by the normalization that relates the geometric
charge to the gauge-theory chemical potential. A quantitative
determination of this response from heavy-ion data will require
reducing this model dependence, for example through a common
calibration of the chemical potential scale across different
holographic backgrounds.

The finite-density trends are results for the chosen charged
background, quadratic dilaton profiles, chemical potential
normalization, and instantaneous freezeout prescription. Their
model dependence under changes of these inputs has not been
systematically assessed here. The instantaneous prescription also
leaves the dynamics of the freezeout stage unresolved.

\section{Acknowledgments}
This work is supported by the National Natural Science Foundation
of China (NSFC) under grant No.~12375140.

\end{document}